\documentclass[prl,floats,aps,showpacs,preprint]{revtex4-1}

\usepackage{graphicx}
\usepackage{dcolumn}
\usepackage{bm}
\usepackage{booktabs}
\usepackage{amsmath}
\usepackage{amssymb}
\usepackage{multirow,hhline,dcolumn}	

\begin{document}

\title{The Stochastic Gravitational Wave Background from the Single-Degenerate Channel of Type Ia Supernovae}

\author{David Falta$^1$}
\author{Robert Fisher$^1$}
\email{robert.fisher@umassd.edu}
\affiliation{$^1$Physics Department, University of Massachusetts Dartmouth, \\ 285 Old Westport Road, North Dartmouth, MA 02747-2300}
\date{\today}

\begin{abstract}
We demonstrate that the integrated gravitational wave signal of Type Ia supernovae (SNe Ia) in the single-degenerate channel out to cosmological distances gives rise to a continuous background to spaceborne gravitational wave detectors, including the Big Bang Observer (BBO) and Deci-Hertz Interferometer Gravitational wave Observatory (DECIGO) planned missions. This gravitational wave background from SNe Ia acts as a noise background in the frequency range 0.1 - 10 Hz, which heretofore was thought to be relatively free from astrophysical sources apart from neutron star binaries, and therefore a key window in which to study primordial gravitational waves generated by inflation.  While inflationary energy scales of $\gtrsim 10^{16}$ GeV yield inflationary gravitational wave backgrounds in excess of our range of predicted backgrounds, for lower energy scales of $\sim10^{15}$ GeV, the inflationary gravitational wave background becomes comparable to the noise background from SNe Ia.
\end{abstract}



\maketitle

\section{Introduction.}
Type Ia supernovae (SNe Ia) are powerful thermonuclear explosions of a white dwarf star that release a kinetic energy of $\sim10^{51}$ ergs, and serve as standard candles for cosmology \cite {phillips93, riessetal98, perlmutteretal99}.  While the precise nature of the SNe Ia progenitor remains a topic of active investigation, current theoretical models   \cite {townsleyetal07, jordanetal09, hoflichetal10} of  the detonation of a white dwarf star brought to a mass near the Chandrasekhar limit by a companion red giant or main sequence star are consistent with both the energetics and nucleosynthetic yield of the explosion, as well as its observed asymmetry \cite {maedaetal10}. We refer to this particular mechanism of SNe Ia as the single degenerate channel (SD)\citep{whelanetal73,nomotoetal84}, as opposed to the double degenerate (DD) channel, which consists of the merger and subsequent detonation of two degenerate white dwarfs \citep {webbink84, ibentutukov84}.

Two models for detonations in the SD channel have been studied extensively -- the deflagration-to-detonation transition (DDT) \citep {khokhlov91, blondinetal11}, and the  gravitationally-confined detonation (GCD). Utilizing detailed three-dimensional (3D) hydrodynamical simulations of the GCD mechanism of the SD channel of SNe Ia, \cite{faltaetal11} demonstrated that the intrinsic asymmetry of the explosion gives rise to a gravitational wave (GW) signature. Based on the GCD model prediction, a nearby individual SNe Ia event may possibly be detected in the Milky Way and nearby galaxies by third-generation GW detectors in the $\sim 2$ Hz frequency range. While the focus of the earlier paper \citep {faltaetal11} was upon the GCD model for detonation in the SD channel,  both the DDT and DD models are likely to yield off-centered detonations \citep {ropkeetal07, vankerkwijketal10}.  Recent work has demonstrated that the explosion energies and nucleosynthetic yields are similar in off-centered detonations in which only a small fraction of the total energy release occurs prior to detonation \citep {chamulaketal11}. Since the great majority of the GW energy is radiated during the detonation itself, we expect both the DDT and DD models are likely to yield off-centered detonations whose gravitational wave signature is qualitatively similar to the GCD model prediction.

Previous work explored the stochastic gravitational wave background due to the distribution of  unresolved ``chirping'' binary white dwarfs, and demonstrated that this background obscures the inflationary signal between frequencies of 10$^{-5}$ Hz and 0.1 Hz \cite {farmerandphinney03}. The frequency range between 0.1 - 10 Hz was previously thought to be relatively free from gravitational wave emission from astrophysical sources, with the primary contribution believed to originate from neutron star binaries \cite {schneider01}, and possibly also core-collapse supernovae at cosmological distances, and Pop III (e.g., first-generation) stars \cite {buonannoetal05}.
In this paper, we consider an additional stochastic GW background source : SNe Ia at cosmological distances. If sufficiently strong, the SNe Ia stochastic background may be a source of noise to future detectors in the 0.1 - 10 Hz frequency range. Thus, the presence of a stochastic gravitational wave background due to SNe Ia may further impose additional challenges in extracting a possible inflationary GW background in this frequency range.

The paper is organized as follows. In section II, we present the formalism underlying the computation of the cosmological gravitational wave background of stochastic sources. In section III, we discuss the cosmic rate of SNe Ia. In section IV, we review the gravitational wave signature of a single SNe Ia event predicted by the GCD mechanism. Finally, in sections V and VI, we present our results and conclude.

\section{Stochastic Background Formalism.}

The GW stochastic background is characterized by the dimensionless ratio of gravitational wave energy density to the critical energy density required to close the universe:
\begin{eqnarray}
\Omega_{GW}(f) = \frac{1}{\rho_c} \frac{d \rho_{GW}}{d \log f} ,
\end{eqnarray}
where $\rho_{GW}$ is the energy density of gravitational waves, $f$ is frequency, and $\rho_c$ is the critical density, given as: 
\begin{eqnarray}
\rho_c = \frac{3 H_0^2}{8\pi G} . 
\end{eqnarray}
Here $G$ is the universal gravitational constant,  and $H_0$ is the Hubble parameter, which we take to be $70.4$ $\mathrm{km}$ $\mathrm{s^{-1}}$ $\mathrm{Mpc^{-1}}$\cite{jarosiketal11}. Using the expression for the energy density of supernovae GW sources, the gravitational wave energy density $\Omega_GW (f)$ can be expressed as an integral over redshift \cite{phinney01}:
\begin{eqnarray}
\Omega_{GW}(f) &=& \frac{8 \pi Gf}{3 c^2 H_0^3 }\int_0^{z}  dz' \frac{R_{SN}(z') }{  E(z')} \frac{d E_{gw}}{df} \Big{|}_{f_r=f(1+z')} ,
\end{eqnarray}
where $c$ is the speed of light, $R_{SN}(z)$ is the SNe Ia event rate per comoving volume, $d E_{gw}/df$ is the GW spectral energy density, and $E(z)$ is a function of both redshift and the cosmological model :
\begin{eqnarray}
E(z) = \sqrt{\Omega_M(1+z)^3 + \Omega_k (1+z)^2 + \Omega_\Lambda} ,
\end{eqnarray}
The spectral energy density is evaluated at the redshifted frequency $f_r=f(1+z)$. We take the currently accepted cosmological parameters consistent with a $\Lambda$-cold-dark-matter model \cite{jarosiketal11}: $\Omega_M=0.27$, $\Omega_\Lambda=0.7$, $\Omega_k=0$.

The duty cycle is an important rate-dependent number that characterizes whether the stochastic background is discrete or continuous. It is determined by
\begin{eqnarray}
D = \int_0^{z} dz' \frac{dR_{SN}}{dz'} V_c(z) \bar{\tau}_{GW} (1+z') .
\end{eqnarray}
Here $\bar{\tau}_{GW}$ is the typical rest-frame duration of the SNe Ia GW signal, which is set by the characteristic detonation time, and $V_c(z)$ is the comoving volume as a function of redshift. If the duty cycle is less than unity, then the signal will consist of a series of rapid bursts separated in time, and relatively easily filtered from other signals, such as the primordial background of gravitational waves generated by inflation.  In contrast, if the duty cycle is unity or larger, the background signal is continuous in time. Such a continuous signal acts as a foreground noise signal to a possible primordial background, and is more challenging to extract from the primordial background. 

\section{The Cosmic SNe Ia Rate.}
The cosmic SNe Ia rate is observationally well-constrained only for low redshifts, $z \lesssim 1$. High-redshift rates are poorly-constrained due to their few-number statistics, but can be estimated using extrapolation and modeling. The SNe Ia rate is modeled as \cite{horiuchiandbeacom10}:
\begin{eqnarray}
R_{Ia}[z(t_c)] = \eta \int_{t_{10}}^{t_z} \Phi(t_c - t_c') \dot{\rho}_*[z(t_c')]dt_c'
\end{eqnarray}
here $\dot{\rho}_*$ the star formation rate (SFR) per unit volume, $\Phi(t)$ is the progenitor-dependent delay time distribution (DTD), $t_{z}$ is the time corresponding to redshift $z$, $t_c$ is the current time, and $\eta$ is the fraction of all stars which ultimately give rise to SNe Ia events. Mathematically, $\eta$ can be written as:
\begin{eqnarray}
\eta = \int_{M_1}^{M_2} A_{Ia}(M) \xi(M) dM ,
\end{eqnarray}
where $\xi(M)$ is the initial mass function of newly formed stars, $A_{Ia}(M)$ is the SN Ia explosion efficiency as a function of initial stellar mass, and the integral is performed over the full range of initial stellar masses $M_1-M_2$ of possible progenitors. If all SNe Ia progenitors exploded promptly at formation, the SNe Ia rate would simply follow the SFR. However because it takes time for a newly formed star to evolve to criticality, the rate is shifted from the SFR by the DTD. Since the SD and DD channels have different ranges of evolution times and relative frequencies that are uncertain, the modeled SNe Ia rate can diverge significantly due to the unknown DTD.

As more high redshift SNe Ia data are collected, the DTD will be further contained and the progenitor model eventually determined. There are currently multiple methods for determining the DTD that use a range of techniques, environments, and redshifts (see \cite{maozetal10} for a review and comparison). Recent studies however generally point to a power-law DTD: $\Phi \propto t^\beta$ with exponent $\beta \approx -1$, suggesting a large component of DD events. Using new data collected from the Subaru Deep Field and previously published rates, \cite{grauretal11} recently found $\beta = -1.1\pm0.1$(statistical) $\pm 0.17$ (systematic).


\section{The Type Ia Gravitational Wave Signal.}

In our previous paper \cite{faltaetal11}, we developed an estimate of the GW signal by post-processing 3D  simulations of the GCD mechanism within the SD channel of SNe Ia \cite {jordanetal08}. We found GW signal with characteristic frequency near 2 Hz, favoring the $h_+$ polarization. The strength of the GCD GW was found to be weak, but detectable as far as 1 Mpc by the Big Bang Observer (BBO), if the event happens to be oriented such that line of sight is close to the equator. In the current paper, we assume that all SNe Ia are generated through edge-lit detonations -- including the GCD, DDT, and DD models -- and   that the GW spectral energy density (figure 4 of \cite{faltaetal11}) of the GCD model is typical of edge-detonated white dwarf (WDs) figure 4 \cite{faltaetal11}. For the current study, in order to accommodate high-redshift supernovae,  we have extended the spectral density to higher frequencies. 

Summarizing our previous SNe Ia GW signal calculation, we simulated the hydrodynamical evolution of C/O WD progenitors from the onset of deflagration through detonation in fully 3D numerical simulations \citep{jordanetal08}. The simulations were advanced within the FLASH framework with the Euler equations of inviscid, non-relativistic hydrodynamics, coupled to Poisson's equation for self-gravity, and an advection-diffusion reaction model of the thickened combustion front. FLASH is a modular, component-based application code framework created to simulate compressible, reactive astrophysical flows and subsequently applied to a wide range of problems \citep{fryxelletal00,dubeyetal09}. These single-bubble GCD models produce intermediate mass elements at a velocity coordinate $\sim 11,000$ km/s, creating a layered structure of IME and Fe peak (NSE) products similar to observation \cite{mazzalietal07, meakinetal09}. However, single-bubble GCD models generally underproduce intermediate mass elements and overproduce $^{56}$Ni. Consequently, the models are generally too luminous in comparison to Branch normal Ia events. 

Our simulation was initialized with a 1.37 $M_\odot$ C/O WD, with a single pre-ignited flame bubble of radius 16 km, and 40 km offset. As the WD evolved, we calculated, for each time step, the second-time derivative of the reduced quadrupole mass moments associated with the standard post-Newtonian expressions for the gravitational quadrupole radiation field from a 3D source \citep{nakamuraetal89}.

\section{Results.}

To constrain the SNe Ia GW stochastic background, we consider an upper-bound SNe Ia rate associated with a prompt DTD model \cite{horiuchiandbeacom10}, which could be characteristic of the shorter evolution of the SD channel, and the more probable $\beta = -1.1$ power-law DTD fit of \cite{grauretal11}.  Here, we extend our integration to $z=10$, which corresponds to the end of reionization,  though our results do not depend sensitively upon this cutoff, as events at much higher redshift contribute negligible power to the overall background. 
 
We estimate the duty cycles associated with our chosen SNe Ia rates, assuming a characteristic detonation time of $\bar{\tau}_{GW}\sim 0.5$ s, obtained from the inverse of the characteristic $2$ Hz detonation frequency \cite{faltaetal11}. Integrating to $z=10$, we find duty cycles that are several orders of magnitude greater than unity for both assumed rates, so that the stochastic background is a continuous signal.



To determine if the SNe Ia GW stochastic background will mask the inflationary signal, we consider a scalar field $\phi$ inflationary model with slow-roll approximation. That is, the potential $V(\phi)$ is assumed to be flat enough that the scalar field rolls slowly and the equations of motion governing inflation can be simplified. We adopt the upper-bound estimate presented by \cite{smithetal06}, constrained by the energy scale of inflation $V^{1/4}$ set by the cosmic background radiation (CMB) and large scale structure (LSS) to be $V^{1/4} \lesssim 3.36 \times 10^{16}$ GeV.

\begin{figure}[H]
\begin{center}
\includegraphics[scale=0.6]{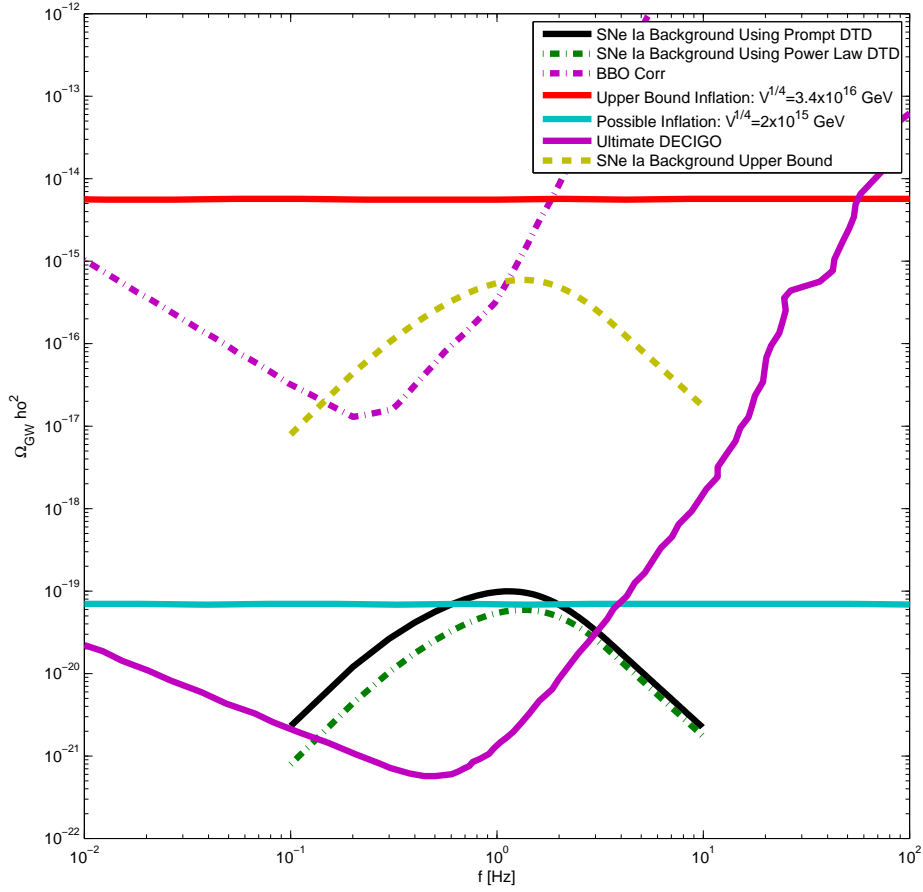}
\end{center}
\caption{Predicted and upper-bound SNe Ia GW stochastic background curves and the proposed correlated BBO and ultimate DECIGO detector sensitivities scaled by the dimensionless Hubble factor $h_0=0.72$. Two possible slow-roll inflationary GW stochastic background curves are presented, an upper-bound estimate derived from the CMB and a second estimate close to the level of the prompt SNe Ia background.}
\label{background_detect}
\end{figure}

In figure \ref{background_detect}, we compare the sensitivities of third-generation detectors and the slow-roll inflation  to our predicted GW stochastic background from cosmological GCD SNe Ia out to $z=10$, assuming all are in the SD channel. Here we also present an upper-bound SNe Ia background estimate, derived by a simple estimate assuming that the SNe Ia event is maximally asymmetric  \cite{faltaetal11}.  This upper-bound signal is obtained by shifting the prompt curve up by a factor of $100^2$, noting that $\Omega_{GW} \propto h^2$ and the upper-bound amplitude is two orders of magnitude larger than our baseline calculated characteristic value.  The baseline calculated value predicts that the stochastic background  could be a source of noise for the inflationary signal for the ultimate form of the correlated Deci-Hertz Interferometer Gravitational wave Observatory (DECIGO) detector \cite{kudohetal06,nakayamaandyokoyama10}. The upper-bound estimate could be a source of noise for even the correlated BBO \cite{buonannoetal05}.


\section{Conclusions.}

A population of purely SD GCD SNe Ia at cosmological distances will generate a continuous GW stochastic background within the 1 - 2 Hz frequency range that could be a source of noise to third-generation detectors seeking the inflationary GW signal. The SNe Ia GW stochastic signal strength is relatively insensitive to the DTD, with the prompt DTD producing a slightly larger signal than the $\beta = -1.1$ power law. The SNe Ia background could be a significant source of noise for the ultimate DECIGO detector if the inflationary signal is at a level less than $\Omega_{GW} h_0 \simeq 10^{-19}$. However assuming a slow roll inflation that can be constrained by the CMB/LSS, the inflationary signal could be as strong as $\Omega_{GW} h_0 \lesssim 6\times 10^{-15}$, in which case the SNe Ia background would be too weak to be a source of noise. If the SNe Ia GW signal strength is significantly stronger, such that the background is closer to the upper bound estimate, it could also be a source of noise to the correlated BBO.


Previous work on the primordial inflationary gravitational wave spectrum has primarily focused upon neutron-star binaries as a source of noise in the frequency range of 0.1 - 10 Hz \cite {schneider01}. The gravitational wave signal generated by these binaries is relatively clean. Therefore it seems that it may be possible to subtract the neutron-star binary spectrum to reveal the underlying primordial gravitational wave spectrum. In contrast, the physics underlying the SNe Ia is far more complex, involving both hydrodynamics and nuclear combustion. Furthermore, though supernova rates are relatively well-constrained from optical observations at low redshift, they are poorly constrained at the high-redshifts responsible for the background.  In this respect, the numerous uncertainties surrounding the stochastic background of SNe Ia are similar to those of core-collapse SNe \cite {buonannoetal05}.  If the inflationary energy scale does lie close to $10^{15}$ GeV, the GW emission of SNe Ia will likely prove to be an additional source of noise to the primordial gravitational wave signal -- one which will be very challenging to extract.



The authors acknowledge useful discussions with Gaurav Khanna. DF acknowledges support from NSF grant numbers PHY-0902026. RF acknowledges research support from NSF grant CNS-0959382 and AFOSR DURIP grant FA9550-10-1-0354. The software used in this work was in part developed by the DOE-supported ASC / Alliance Flash Center at the University of Chicago. This research was supported in part by the National Science Foundation through TeraGrid resources provided by the  Louisiana Optical Network Initiative under grant number TG-AST100038.

\newcommand{\aj}{Astronomical Journal}
\newcommand{\apjl}{Astrophysical Journal Letters}
\newcommand{\apjs}{Astrophysical Journal Supplement}
\newcommand{\aap}{Astronomy and Astrophysics}
\newcommand{\araa}{Annual Review of Astronomy and Astrophysics}
\newcommand{\mnras}{Monthly Notices of the Royal Astronomical Society}
\newcommand{\jcap}{Journal of Cosmology and Astroparticle Physics}


\onecolumngrid

\end{document}